 \documentclass[preprint2]{aastex}

\usepackage{xspace}
\usepackage{color}

\newcommand{\msun}{M$_{\odot}$}

\newcommand{\sbs}{SBS\,1150+599A}
\newcommand{\png}{PN\,G\,135.9+55.9\xspace}

\newcommand{\ergsa}{~erg~s$^{-1}$~cm$^{-2}$~\AA$^{-1}$}

\newcommand{\nhe}{n_{\mathrm{He}}/n_{\mathrm{H}}}

\newcommand{\Teff}{\ensuremath{T_{\rm{eff}}}}

\newcommand{\Hb}{\ifmmode {\rm H}\beta \else H$\beta$\fi}
\newcommand{\hi}{H~{\sc i}}

\slugcomment{}

\shorttitle{Binary Core of Planetary Nebula PN\,G\,135.9+55.9}
\shortauthors{ Tovmassian et al.}

\begin{document}
\title{A close binary nucleus in the most oxygen-poor planetary nebula
        PN\,G\,135.9+55.9.}


\author{G.H. Tovmassian \altaffilmark{ }}
\affil{Observatorio Astron\'omico Nacional, Instituto de  Astronom\'\i a
, UNAM, \\ P.O. Box 439027, San Diego,
                      CA 92143-9027, USA}
\email{gag@astrosen.unam.mx}

\author{R. Napiwotzki\altaffilmark{ }}
\affil{Department of Physics \& Astronomy,\\ University of Leicester,
University Road, Leicester LE1 7RH, UK}
\email{rn38@astro.le.ac.uk}

\author{M. G. Richer \altaffilmark{*}}
\affil{Observatorio Astron\'omico Nacional, Instituto de   Astronom\'\i a
, UNAM,\\ P.O. Box 439027, San Diego,
                      CA 92143-9027, USA}
\email{richer@astrosen.unam.mx}

\author{G. Stasi\'nska \altaffilmark{*}}
\affil{LUTH, Observatoire de Meudon, 5 Place Jules Janssen,
                      F-92195 Meudon Cedex, France}
\email{grazyna.stasinska@obspm.fr}

  \author{A.W. Fullerton\altaffilmark{**}}
\affil{Deptartment of Physics and Astronomy, University of
Victoria,\\ P.O. Box 3055, Victoria, BC V8W 3P6,
                     Canada}
\email{awf@pha.jhu.edu}
\vspace{-3cm}
\and

\author{T. Rauch}
\affil{
  Institut f\"ur Astronomie und Astrophysik T\"{u}bingen (IAAT)
  Abteilung Astronomie \\
  Sand 1
D-72076, T\"{u}bingen,
Germany }
\email{rauch@astro.uni-tuebingen.de}


\altaffiltext{*}{Visiting Astronomer, CFHT}
\altaffiltext{**}{Center for Astrophysical Sciences, Department of
Physics and Astronomy, Johns Hopkins University,
                     3400 North Charles Street, Baltimore, MD 21286, USA}


\begin{abstract}
We report FUSE\footnote{Based on observations made with the
NASA-CNES-CSA Far Ultraviolet Spectroscopic Explorer. FUSE is
operated for NASA by the Johns Hopkins University under NASA
contract NAS5-32985.}  and new deep optical spectroscopic
observations of \png, the most oxygen-poor planetary nebula
located in the Galactic halo. These observations allow us to
estimate the gravity of the central star by fitting the profile of
the observed H~{\sc i} absorption lines with NLTE model
atmospheres.  Our best fit implies that the central star is still
in a pre-white dwarf stage.  We also find large variability of the
radial velocities of the absorption component of the Balmer lines
on a timescale of hours. This is direct evidence that the nucleus
of \png\ is a close binary. The large semi-amplitude of the radial
velocity variations and the probably short period suggest a
massive compact companion, likely a white dwarf. Although our
orbital solutions are very preliminary, they indicate that the
total mass of the system probably exceeds the Chandrasekhar limit.
If confirmed, this would make this binary the potential progenitor
of a type Ia supernova.

\end{abstract}





\section{Introduction}

\png\ (also called \sbs) is a recently discovered planetary nebula
in the Galactic halo \citep{tovmassianetal2001} whose oxygen
abundance is reported to be extremely low, of order 1/100 of the
solar value or less
\citep{tovmassianetal2001,richeretal2002,jacobyetal2002}. This
makes it, by one order of magnitude, the most oxygen-poor
planetary nebula known so far. Such a low oxygen abundance may be
either genuine and represent the oxygen abundance in the material
from which the progenitor star formed, or it may be the result of
mixing processes while the star was on the AGB. Although the
presently available determinations of the Ne/O ratio are
discrepant by a factor of about 10 (Richer et al. 2002 and Jacoby
et al. 2002), they indicate that, even in the case of conversion
of O into Ne, the original oxygen abundance of the progenitor of
\png\ is likely extremely low.

Clearly, this unique object deserves a more detailed study to
better define its nature. Tovmassian et al. (2001) speculated upon
various scenarios for the formation of this planetary nebula. One
of them assumed a progenitor star having formed out of infalling
material. Another invoked the existence of a close binary core. In
this paper we focus on the information provided by new
observations concerning the nature of the central star of \png. A
companion paper (Stasi\'nska et al. in preparation) will discuss
new observational constraints on the chemical composition of this
object.

First, we acquired far-ultraviolet spectra of \png\ using the Far
Ultraviolet Spectroscopic Explorer (FUSE). We also obtained a
series of high signal-to-noise optical spectra of the nebula and
its central star. From these observations, one can constrain the
fundamental parameters of the central star. It is also possible to
investigate whether there might be variability on short time
scales. The most surprising result of our study is the discovery
that the core of this planetary nebula is a short period,
close binary with an unseen companion, probably a white dwarf.

The paper is organized as follows: In Section 2, we describe the
observations and the reduction procedure, both for the FUSE and
ground-based observations. In Section 3, we describe the main
spectral features of the object from the FUSE and ground-based
observations. In Section 4, we present a stellar atmosphere
analysis that allows us to infer the gravity of the ionizing star
from the observed H~{\sc i} absorption features and estimate
parameters of the binary system. In Section 5, we present
estimates of the reddening of \png\ based upon the FUSE data and
in Section 6 we estimate the distance and age of \png.  In Section
7, we discuss the binary nature of the core and the masses of its
components. Finally, in Section 8, we summarize the main results
of this study and present some prospects for future study.

\section{Observations}
\subsection{FUSE observations}

FUSE is an orbiting observatory with the capability to obtain high
resolution spectroscopy in the wavelength region between 900 and
1200\,\AA. It consists of four co-aligned telescopes optimized for
far-UV wavelengths. The spherical, aberration-corrected,
holographic diffraction gratings disperse light from four
channels. Two channels with SiC coatings cover the 905--1100\AA\
region and two others with Al+LiF coatings detect the
1000--1187\AA\ range. An overview of the FUSE mission has been
given by \citet{moos}.

The observation  of \png\  (ID\# C034) was conducted through the
LWRS aperture ($30''\times30''$ square) on 2002 January 30 for a
total exposure time of 30\,660\,s. The observation was composed of
seven individual integrations of approximately 4200\,s each,
simultaneous in all channels. These observations were among the
first executed after the recovery from the failure of two reaction
wheels in December 2001. Therefore, the pointing was not as good
as was achieved before the accident or after the situation was
taken under complete control. We have checked possible deviations
during the observations. There was a systematic, quasi-sinusoidal
pointing drift on an orbital time-scale. However, the maximum
displacement is only 2\farcs3, which corresponds to a maximum
shift of 0.026 \AA\ (i.e., 7.7 km/s at 1000 \AA), or about half a
spectral resolution element. Thus, it can hardly pose a problem,
but the pointing itself (centering in the aperture), which is
difficult to verify, can introduce zero-point offsets of up to
0.15\AA. The pipeline reduction package CalFUSE 2.1.7 was used for
the extraction of the spectra as well as reduction and calibration
purposes.

Our examination of the spectra showed that the wavelength shifts
between  individual exposures, which are common for FUSE data,
never exceeded 1.5 pixels and were usually in the 0.1-0.5 pixel
range. Since the shifts were negligible, we used a non-standard
processing wherein all 7 exposures were concatenated into one
large photon list before pushing it through CalFUSE 2.1.7.  This
permits a more reliable background model to be created, which is
especially important for faint targets such as \png, and also
eliminates the need to cross-correlate and coadd exposures. The
other important procedure that we used was to separate the night
exposure (NE) from the total exposure of the object. The data are
severely contaminated by the airglow lines arising form the
illuminated atmosphere of the Earth during day time
\citep{feldman}. In order  to eliminate the strong emission lines
from the terrestrial airglow, we extracted photons obtained when
the exposure was taken during orbital night. This decreased our
exposure time by factor of 3. We use both the total exposure and
NE spectra to analyze the data of \png. No event bursts or other
unusually strong glitches common for FUSE data were detected in
the raw images or the photon arrival rate plots  \citep[see The
FUSE Instrument and Data Handbook edited by ][]{fusehandbook}.

\subsection{CFHT optical observations}

The observations at the Canada-France-Hawaii Telescope (CFHT) were
obtained on 2003 May 1 UT with the MOS spectrograph
\citep{lefevreetal94}. The U900 grism was used to observe the
3400--5300\,\AA\ spectral interval.  A $1\arcsec$ slit was used.
Coupled with the EEV1 CCD, $2048\times 3900\times 13.5 \mu$m
pixels, the dispersion was 0.78\,\AA/pix and the resolution
3.0--3.5\,\AA, based upon the widths of arc lamp lines. A $B$
filter (CFHT filter \#1412) was used to center the central star of
\png\ in the slit.  The spectra of \png\ were obtained in pairs
with the spectrograph slit set at the parallactic angle
corresponding to the midpoint of the pair of spectra.  This was
done since the spectrograph rotation and object re-acquisition
took a significant amount of time. For the observations of \png\
and the standard star BD+33$^\circ$2642 \citep{oke1990}, the slit
position angle was within $10^\circ$ of the parallactic angle at
all times. Seven 1800\,s exposures of \png\ were obtained spanning
the airmass range 1.30--2.65.  The observations of
BD+33$^\circ$2642 were obtained using a $5\arcsec$ slit. Spectra
of Hg, Ne, and Ar lamps were obtained for wavelength calibration
at the end of the night only. Images of the flat field lamp were
obtained through both slits to correct for pixel-to-pixel
variations.  Bias images were also obtained in order to remove any
two-dimensional structure in the bias level.

The data were reduced using the Image Reduction and Analysis
Facility (IRAF)\footnote{IRAF is distributed by the National
Optical Astronomical Observatories, which is operated by the
Associated Universities for Research in Astronomy, Inc., under
contract to the National Science Foundation.} software package
(specifically the specred package).  The mean level of the
overscan region was subtracted from each image.  Next, the average
of the overscan-corrected bias images was subtracted from all
object and flat field images to remove any two-dimensional bias
structure. The flat field images were combined and then processed
to divide out the shape of the flat field lamp.  The resulting
flat field was then divided into all object images.  The spectra
of \png\ and the standard star were extracted to one-dimensional
images, subtracting the sky contribution by defining sky apertures
on both sides of the object spectra and interpolating between them
with a straight line.  The wavelength axis was calibrated using
the spectra of the arc lamp.  Since the object was observed
repeatedly over a large airmass interval, an empirical atmospheric
extinction curve was derived from the data. This extinction curve
turned out to be somewhat steeper than that given in the CFHT
Observatory Manual\footnote{See
http://www.cfht.hawaii.edu/Instruments/ObservatoryManual/}.  The
spectra were calibrated in flux using the observations of the
standard star and the previous empirical atmospheric extinction
curve.  The individual spectra of \png\ were completely calibrated
before combining or further analyzing them.

\section{The main features of the observed spectra}

\begin{figure*}
\includegraphics[width=166mm, bb=15 145 570 700, clip]{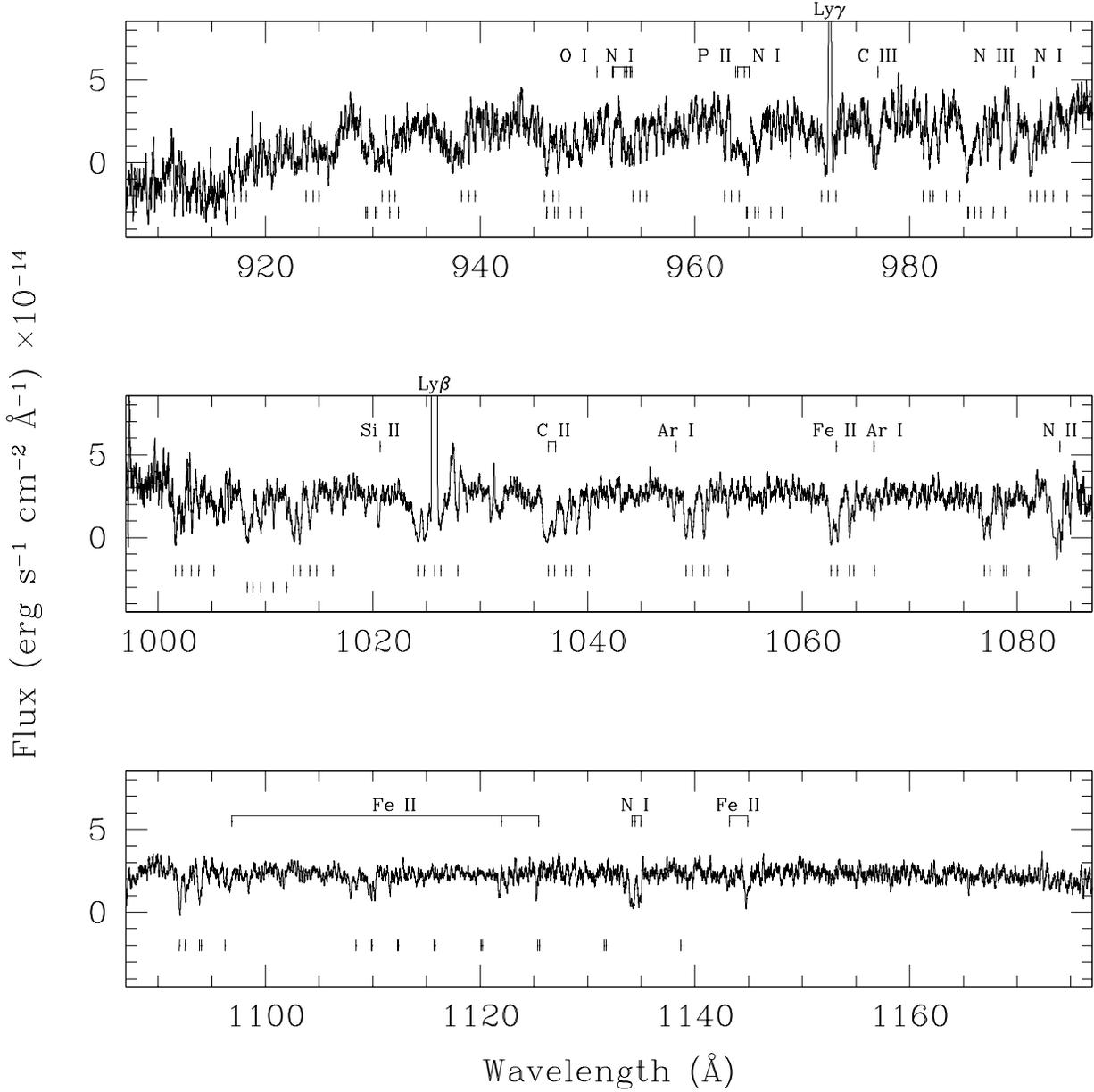}
\caption{The FUSE composite spectrum of \png\ for the night only
exposure. A few of the strongest H\,{\sc i} and O\,{\sc i} $\lambda
1025.76$\,\AA\  airglow lines remain in emission. The other features are
absorption lines, most of which are due to interstellar matter.
Molecular hydrogen lines are marked along the bottom of the spectrum:
the tick marks for the Lyman series lines are above those for the
Werner series. Some other prominent interstellar lines of other
elements are marked above the spectrum. The rest positions of Ly$\beta$
and Ly$\gamma$ are marked above the corresponding boxes. } \label{spec}
\end{figure*}

The FUSE spectrum itself is an approximately flat continuum, with
a mean level of about $3\times10^{-14}$\ergsa.  There are many
lines of interstellar (IS) H$_2$. No emission lines could be seen
except those that are definitely associated with terrestrial
airglow. Most of these lines disappear in the NE spectrum, with
only the strongest lines of \hi\ and O~{\sc i} leaving residual
emission peaks whose intensities are dramatically reduced compared
to the all photon spectrum. Since the systemic velocity of \png\
is $-$193 km/sec \citep{tovmassianetal2001,richeretal2003}, there
is no possible confusion between airglow emission lines and lines
from the object. The FUSE NE spectrum of \png\ is presented in Fig
\ref{spec}. On the whole, it is impressively featureless,
particularly when compared with a large subset of FUV spectra of
planetary nebulae from the FUSE archives that we examined. This
remarkable absence of features in the FUSE spectrum of \png\ is
consistent with the very low metal abundances derived from the
analysis of the PN.

\subsection{Absorption lines from the interstellar medium in the FUSE spectrum}

Most of the absorption lines that are observed originate in the
interstellar matter between us and the object.  There are numerous
lines of molecular and atomic hydrogen. Their study can provide
interesting constraints upon the distribution of H and H$_2$ along
the line of sight toward the object \citep[see
e.g.][]{McCandliss03}. However, we are mainly concerned with the
physics of \png\ itself in this paper and discuss only the impact
of these interstellar lines upon our analysis of \png.

Molecular hydrogen is a significant source of contamination when
studying the spectral features of the underlying source.
Fortunately, the physics of absorption by molecular hydrogen is
well understood and can be reliably modeled. Absorption bands of
interstellar molecular hydrogen arise from the photo-excitation of
molecular hydrogen by the background continuum object, causing
electrons in the ground electronic state to be excited into higher
states of either the Lyman or Werner bands. The strength of these
absorption lines is proportional to the molecular column density
distribution within the ground electronic state's rotational
($J''$) and vibrational $(v'')^2$ energy levels along a particular
line-of-sight. We used the FUSE Simulator Tool to model the FUSE
spectrum ({\sl http://violet.pha.jhu.edu/\~\,gak/fwebsim.html}).
In order to estimate the hydrogen column density, we used a flat
spectrum of intensity $3\times10^{-14}$\ergsa\ and a variety of
parameters for the foreground column densities of atomic and
molecular hydrogen. Later, we repeated this analysis with  our
derived model spectrum, described in the following sections, and
obtained exactly the same results.  A grid of values of neutral
hydrogen column densities ranging from log $N_{\mathrm{H}}$ = 20
to $20.6\,\mathrm {cm^{-2}}$ were investigated. For H$_ 2$, the
input values were log N($v'', J''$) = 18 to 20. The model uses
Voigt profiles to calculate the optical depth variation with
wavelength for each line. Line shapes reproducing the linear,
flat, and square root portions of the curve-of-growth are
controlled through the specification of the Doppler parameter $b$.
The Doppler parameter $b$ of line width was adopted equal to 10
km~s$^{-1}$, as provided by default in the simulation tool.

The modelled interstellar absorption shows a significant 0.2\AA\
wavelength shift compared to the observed spectrum as estimated
from cross-correlation. This shift is larger than expected for
interstellar hydrogen absorption, e.g., \cite{bannister} did not
measure velocities in excess of 30\,km/s for interstellar lines
from atomic species in a sample of white dwarfs with high
resolution HST UV spectra. Even considering the much larger
distance to \png\ compared to that sample, it is unlikely that the
entire wavelength shift is due to high velocity clouds along this
line of sight.  At least half of the shift in the FUSE spectrum of
\png\ (0.2 \AA\ corresponds to approximately 60 km/sec) is
probably caused by imperfect centering of the central star in the
LWRS aperture of FUSE. This is a well-known problem that can cause
shifts of the wavelength scale. We adopted the view that the
radial velocity shift of the H$_2$ features is caused by this
instrumental effect and corrected the wavelength scale for further
fitting of the photospheric line profile.

\begin{figure*}
{\includegraphics[width=83mm, bb=20 150 570 700, clip]{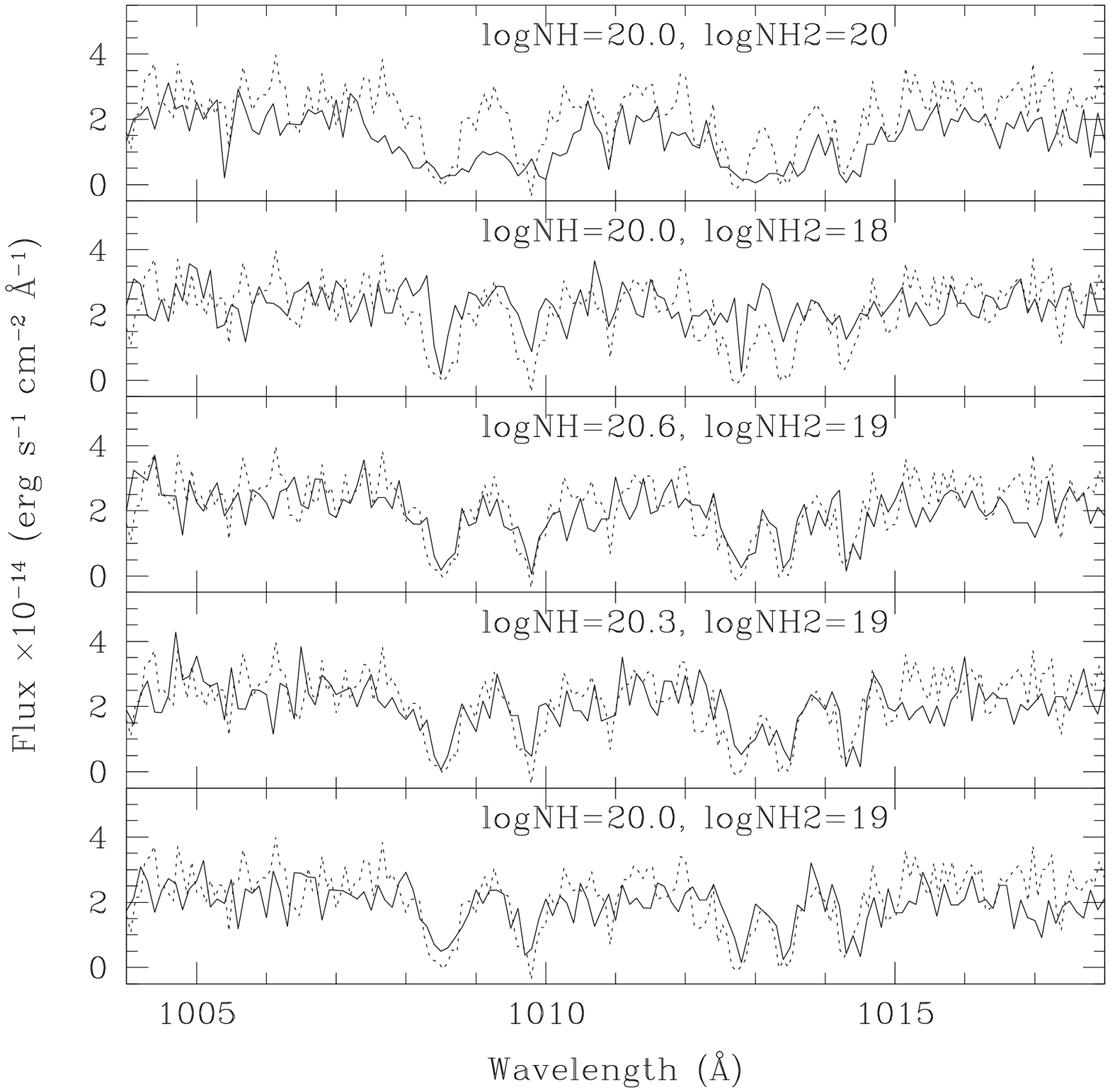}}
{\includegraphics[width=83mm, bb=40 150 590 700, clip]{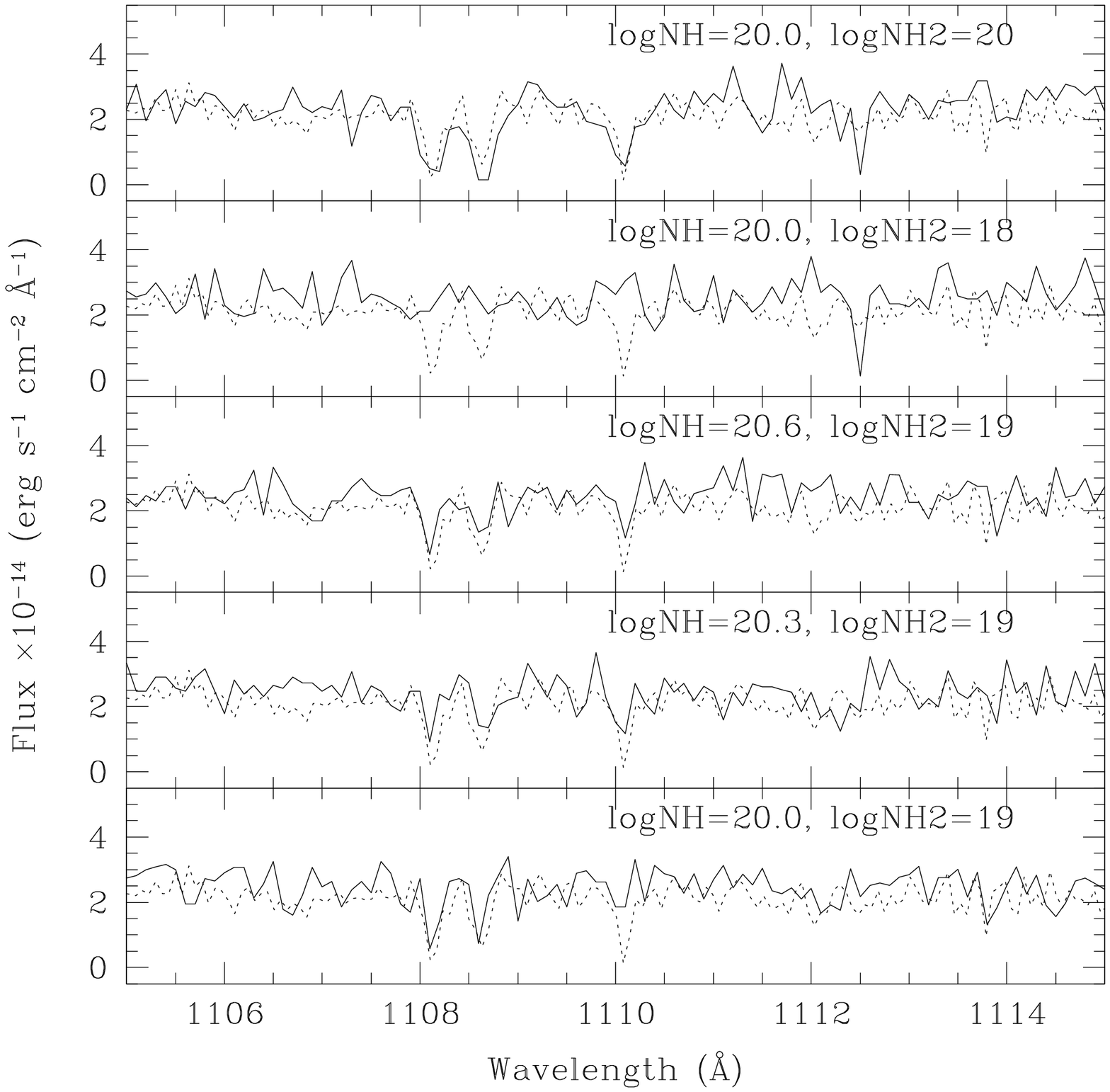}}
\caption{Two portions of all photon FUSE spectrum (LiF 1A and 1B
channels) of \png\ that include prominent interstellar lines of
H$_2$ are plotted with a dotted line. Overplotted are modelled
spectra that assume a flat underlying spectrum of intensity
$3\times10^{-14}$\ergsa\ and different column densities of neutral
and molecular hydrogen.} \label{h21}
\end{figure*}

The H$_2$ column density strongly affects the profiles of the
H$_2$ lines, whereas the impact of the neutral hydrogen column
density is less significant. In Fig.~\ref{h21} simulated spectra
with different hydrogen (H\,{\sc i}, H$_2$) column densities are
presented. Two spectral regions are shown.  The first is around
$\lambda\,1012$\,\AA, where the Werner series has strong lines,
while the other is around $\lambda\,1108$\,\AA, with prominent
H$_2$ lines of the Lyman series. In both regions, even without
doing a $\chi^2$ fit, it is obvious that values of log
$N_{\mathrm{H_2}}$ of 18 and 20 dex are far from fitting the
observed line profiles and intensities, while $\log
N_{\mathrm{H_2}} = 19$ dex provides a good fit. For neutral
hydrogen, the choice of the best fit is not so obvious. Another
useful web tool allows one to calculate the column density
depending upon the direction in the Galaxy based upon the results
of  \citet[][{\sl
http://heasarc.gsfc.nasa.gov/cgi-bin/Tools/w3nh/w3nh.pl}]{dilo90}.
According to this calculation, the expected neutral hydrogen
column density is $N_{\mathrm{H}} \approx
1.53\times10^{20}\,\mathrm {cm^{-2}}$ in the direction of \png.
This roughly corresponds to the modelled value of log
$N_{\mathrm{H}}$ = 20.3, which is shown in the second panel from
the bottom in Fig.~\ref{h21}. However a higher value of $\log
N_{\mathrm{H_2}} = 20.6$ seems to be an equally good fit. We
suspect that the column density toward \png\ may be slightly
higher than 20.3 dex, as follows also from the analysis of
absorbed flux in the UV. This will be discussed in detail below in
Sect. 5.

Besides the numerous H$_2$ lines of the Werner and Lyman series, there
is a large number of absorption lines from O~{\sc i}, N~{\sc i}, N~{\sc
ii}, N~{\sc iii} and  Fe~{\sc ii} that are also due to the intervening
interstellar medium.
These lines are commonly encountered during FUSE observations
   \citep[e.g.][]{fro2001,wo2001} and, most
importantly, their wavelengths indicate that they are not related
to our object. The fact that these lines are numerous and strong
is in agreement with the previously deduced large distance to the
object and its location in the Galactic Halo \citep[][see also
Sect. 6]{tovmassianetal2001,richeretal2002}.

\subsection{Hydrogen and helium features physically related to \png}

The only spectral feature in the FUSE spectrum firmly identified
to arise in our object, apart from the continuum, is the
relatively broad absorption from Ly$\beta$ that we attribute to
the central star. The only possible emission feature from the
nebula itself is a small emission peak inside the Ly$\beta$
absorption. As mentioned above, we expect the lines from the
central star to be Doppler shifted to the blue by the heliocentric
systemic velocity of the nebula,  -193.3 km s$^{-1}$
\citep{richeretal2003}. The nebular lines may appear to be
blue-shifted somewhat more, to as much as -205 km s$^{-1}$,
because  the nebular emission consists of two components with the
blue-shifted component being the stronger of the two
\citep{richeretal2003}.  The portion of the all photon spectrum
around Ly$\beta$ is presented in Fig.~\ref{f:Lyfit} along with the
models that were fit to it (described below).

The profile of the absorption feature blueward of the vacuum
wavelength position of Ly$\beta$, indicated by a strong airglow
line, is extremely complicated. First, the absorption line is
saturated.  Second, the red wing is contaminated by the airglow
emission whose profile itself is complicated by a small glitch on
the blue wing. Third, there is residual emission from the O~{\sc
i} airglow line further to the red. Fourth, there are several
strong components of interstellar H$_2$ Lyman lines inside the
stellar absorption line. Finally, inside the absorption feature
sits an emission peak that we interpret as nebular He~{\sc ii} 6-2
$\lambda 1025.273$ emission (see below) superposed upon the H~{\sc
i} Ly$\beta$ absorption feature from the central star. The
emission line that we identify as He~{\sc ii} has a measured
wavelength of 1024.57 \AA. This coincides exactly with the
position of He~{\sc ii} $\lambda 1025.273$ at a radial velocity of
-205 km s$^{-1}$.

We corrected the line profile in the all photon spectrum, applying
the model of interstellar absorption with the parameters estimated
in the previous section (log $N_{\mathrm{H}}$ = 20.3 and
$N_{\mathrm{H_2}}$ = 19). The residual spectrum still shows broad
 photospheric Ly$\beta$ absorption as well as emission at the position of
He~{\sc ii} line.  Although the strength of the absorption feature
corresponding to Ly$\beta$ depends upon the parameters adopted for
interstellar hydrogen in the model, it is both expected and
present in the spectrum. We are less confident regarding the
He~{\sc ii} emission feature, because it is located in a gap
between two strong interstellar lines and can be easily confused
with the residual intervening continuum. However this supposed
emission line protrudes more than can be accounted for by the
models, even if column densities exceeding $\log N_{\mathrm{H}} =
20.3$ are considered. If this is the case, then the upper limit to
the flux from the He~{\sc ii} $\lambda 1025.273$ is $\leq
4.4\times10^{-16}${~erg~cm$^{-2}$~s$^{-1}$}, according to the
model fit that is described in Sect. 4.2.

In the optical range, we discovered signatures of the central
object in the form of absorption lines at the wavelengths of
hydrogen and helium lines. These features were not previously
noted because our new spectra have higher S/N (by a factor of 3)
and higher spectral resolution than all of our previous spectra.
As an experiment, we degraded our new CFHT spectra to the
resolution that we used before, and the absorption features almost
disappeared, leaving marginal traces only in higher members of the
Balmer series.  Our previous observations also used a wider slit,
substantially increasing the nebular contribution and possibly
widening the emission lines.

As detailed in the Table~\ref{rvtbl}, we obtained seven optical
spectra of \png\ during one night. The observations were primarily
designed to obtain deep spectrophotometric data in far optical UV
for study of Ne ions located there and other possible faint
emission lines from the nebula in the integrated spectrum. These
will be discussed in a subsequent publication concerning the
nebular characteristics. Here we concentrate on the features due to
the central star. Even a simple eye inspection of the co-added
spectrum immediately attracted our attention to the dips around
higher Balmer series lines, which are clearly not centered on the
emission lines. An analysis of the individual spectra revealed
that the center of the absorption component varies, indicating
large radial velocity shifts.

In principle, our observing technique and equipment are not ideal
for detecting radial velocity variations.  The use of a grism
spectrograph, with a nonlinear dispersion, the large range of
telescope inclination angles spanned by the observations, and the
variable slit angle combined to scatter the absolute zero point of
the wavelength scale among the spectra by up to 10\,\AA. However,
all of our measurements of absorption components in the central
star spectrum that are relevant to this study where made relative
to corresponding nebular emission line components, which
effectively eliminates possible problems.

\section{Quantative analysis of the stellar core of \png}

\subsection{First determinations of the radial velocity variations}
\label{s:first}

Initial estimates of the radial velocities (RVs) were obtained
by de-blending the line profiles with the fit routines of the
gnuplot program.  A Gaussian profile was assumed for the emission
component (from the nebula). A Voigt profile was taken for the
absorption component, as it is better suited to fit a broad white
dwarf-like absorption component. We succeeded in separating the
components for the lines where we had enough signal, enough
continuum around the lines, and an absorption component
sufficiently strong to be resolved. This concerns primarily
H$\gamma$, H$\delta$, and H$\epsilon$, where de-blending worked
fine for all seven spectra (i.e. for different orbital phases),
returning consistent parameters of the measured lines from one
line to another and from spectrum to spectrum. To reduce the
number of free parameters, we performed a simultaneous fit of all
spectra. The FWHM of the Gaussian used to fit the emission lines
and the parameters of the Lorentzian fit to the photospheric
absorption were held fixed, because these values should remain
constant, while all other parameters were allowed to vary from
spectrum to spectrum.

The imperfections of wavelength calibration were eliminated by
measuring the radial velocities of the absorption components
relative to their emission counterparts, which we suppose to be
constant in general, and on such short time scale in particular,
as is also evident from our previous observations
\citep{richeretal2002}. The corresponding measurements of the
absorption components along with epochs of observation are
presented in the second and sixth columns of Table~\ref{rvtbl}.
Clearly, the radial velocity variations of the absorption
components of these three  lines are consistent with each other,
and vary strongly on a timescale of a few hours. Before analyzing
their time dependance, i.e. periodic nature, let us consider
atmospheric models of the central star and estimate radial
velocities from the absorption line fits.

\begin{table*}
\caption{RV measurements from the optical spectra. The
heliocentric Julian date of the center of the exposures and the
RV shifts relative to the nebular emission lines are given\tablenotemark{b}.}
\vspace{3mm}
\label{rvtbl}
\begin{tabular}{ccrrrr}
\tableline\tableline
spectrum &HJD$-2,450,000$ &RV  atm. model fit /&
   \multicolumn{3}{c}{ RV  de-blending} \\
number   &                &5 lines composite & H$\gamma$ &
H$\delta$ & H$\epsilon$ \\  \tableline
457    &2760.80249 & $-204\pm 25$ & -95 & -136 &  -180 \\
466    &2760.90086 & $269\pm 28$  & 257 & 236  &  365  \\
467    &2760.92412 & $179\pm 25$  & 208 &  220 & 258 \\
471    &2760.95986 & $-53\pm 60$     & 23  & -7  & 0 \\
472    &2760.98410 & $-59\pm 60$     & 28 & 16  & -36 \\
476    &2761.01893 & $111\pm 28$  & 178 & 138  & 164 \\
477    &2761.04169 & $246\pm 28$   & 229 & 236  & 264 \\
\tableline
 \end{tabular}
\tablenotetext{b}{RV errors estimated from measurements of
emission line component does not exceed $\pm 10$ km/sec.}
\end{table*}

\subsection{Model atmosphere calculations}
\label{atmospheres}

The model atmosphere spectra used for our analysis were computed
with the NLTE code PRO\,2 developed by  \citet{wer86}. The basic
assumptions are those of static, plane-parallel atmospheres in
hydrostatic and radiative equilibrium. As described in
\citet{wer86} and \citet{WD1999}, the accelerated lambda iteration
(ALI) method is used to solve the set of non-linear equations.

The basic parameters, effective temperature and surface gravity, were
determined with the grid of model atmospheres composed of
hydrogen and helium calculated by  \citet{napw1999} for the
analysis of planetary nebulae central stars. Extensive model
atoms for hydrogen and helium (\ion{He}{1} and \ion{He}{2}) were
used \citep[cf.][for details]{napw1999}. Line profiles
of the hydrogen Balmer and Lyman lines and the \ion{He}{2} lines were
computed using extended VCS broadening
tables \citep{vid1973} provided by  \citet{lem97} and
  \citet{schbu89}, respectively.

\subsection{The gravity of the central star from model atmosphere analysis}

It has now become a standard technique to determine both the
temperature and gravity of central stars of planetary nebulae or
of hot white dwarfs from a simultaneous fit of the Balmer series
\citep[e.g.][]{napw1999} or Lyman series \citep{bartsowetal}.

In our FUSE spectrum, only the Ly\,$\beta$
line has a reasonable signal-to-noise. Higher
members of the Lyman series fall in the SiC channels, which are
characterized by lower effective area and hence lower S/N. However, the
analysis of the Ly\,$\beta$ line is problematic as well, because major
parts of the line profile are contaminated by geocoronal emission or
absorption caused by interstellar hydrogen molecules.

The best opportunity for parameter estimation is offered by the
hydrogen Balmer lines in the optical spectrum, although these lines
are also contaminated by the strong emission
coming from the nebula.  The emission subsides toward higher members
of the Balmer series, while the decline of the photospheric absorption lines is less
pronounced.
Thus, it was possible to fit five Balmer lines (from
H$\gamma$ to H$_9$).

The model atmosphere analysis was done with the programme {\sc
fitsb2}, which was originally developed to analyze the spectra of
double-lined white dwarfs from the SPY project
\citet{NCD2003,NYN2004}. However, this programme is also
well-suited for the analysis of the spectra of single-lined
binaries. {\sc fitsb2} performs a simultaneous fit of spectra
covering different orbital phases, i.e., all available information
is combined into the parameter determination procedure. Fit
results are stellar parameters and RVs \citep[a more detailed
discussion is given in][]{NYN2004}.  Two \ion{He}{2} lines
(4540\,\AA\ and 4686\,\AA) are visible as well and enabled us to
estimate the helium abundance.

We used the nebular emission lines to correct the wavelength
calibration as described in Sect.~\ref{s:first}. Since the emission
components of the higher Balmer lines are weak, individual
wavelength measurements become less accurate. Thus we performed an
average over all spectra and calculated wavelength corrections from
a simple linear fit and applied this for all lines higher than
H$\gamma$. Deviations of individual measurements for H$\gamma$,
H$\delta$, and H$\epsilon$, which are most important for the RV
determination, never exceeded 10\,km/s.

Due to the nebular contamination of the available spectra, it was
not possible to determine both the temperature and gravity from
the available spectra. Thus, we kept the temperature fixed during
the fitting procedure and determined only the surface gravity. We
considered \Teff\ values in the range 80000\,K to 150000\,K. The
results are listed in Table~\ref{tgfit}. Due to potential problems
with degradation of the line profile by orbital smearing (cf.\
discussion below), we restricted the model atmosphere fit to the
three spectra with the largest RV shifts (shown in Fig.~3). These
were probably taken close to quadrature phases, for which the rate
of change of RV and thus the orbital smearing is smallest.

\begin{figure*}
\includegraphics[angle=-90,width=160mm]{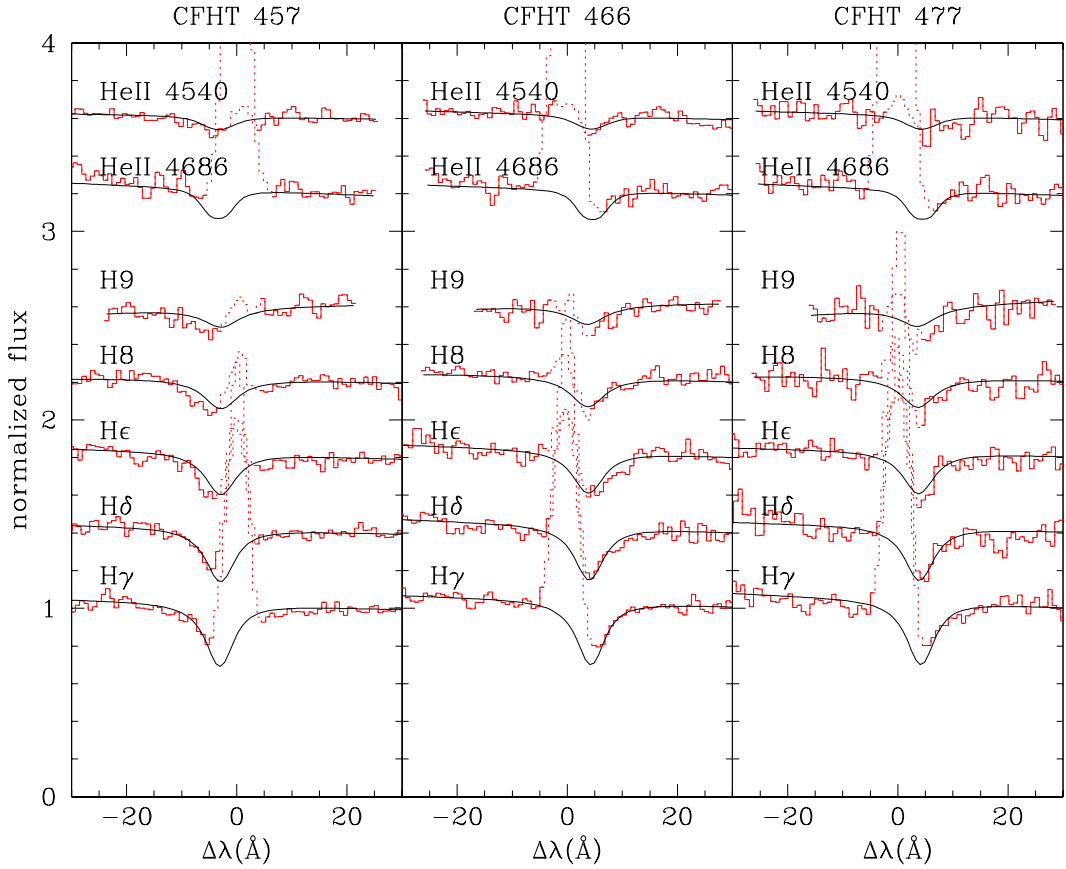}
\caption{A model atmosphere fit (solid line) to the optical
spectra (histogram) taken during the phases with largest RV shifts.
The dotted lines indicate parts of the spectra which were excluded from the
fit.} \label{balmerfit}
\end{figure*}

\begin{table}
\caption{Results of the model atmosphere fits of the quadrature
  spectra 457, 466, and 477 for several values of \Teff.
Masses and
  radii were derived from an interpolation in the post-AGB tracks in
  Fig.~\ref{tefflogg}.}
\vspace{3mm}
\label{tgfit}
\begin{tabular}{rrrrr}
\tableline\tableline\\ \Teff & $\log g$ & $\log(\nhe)$
         &$M/M_\odot$   &$R/R_\odot$\\
   K &  $\mathrm{cm\ s^{-2}}$ & &\\     \hline
80000   & 4.99     &               &0.62  &0.42\\
90000   & 5.05     &               &0.68  &0.41\\
100000  & 5.15     &               &0.75  &0.38\\
110000  & 5.26     &               &0.83  &0.35\\
120000  & 5.35     & -1.00         &0.88  &0.33\\
130000  & 5.39     &               &0.96  &0.33\\
150000  & 5.56     &               &1.02  &0.28\\
\tableline\tableline
\end{tabular}
\label{atmmod}
\end{table}

\begin{figure}[h]
\includegraphics[width=55mm, angle=-90, clip]{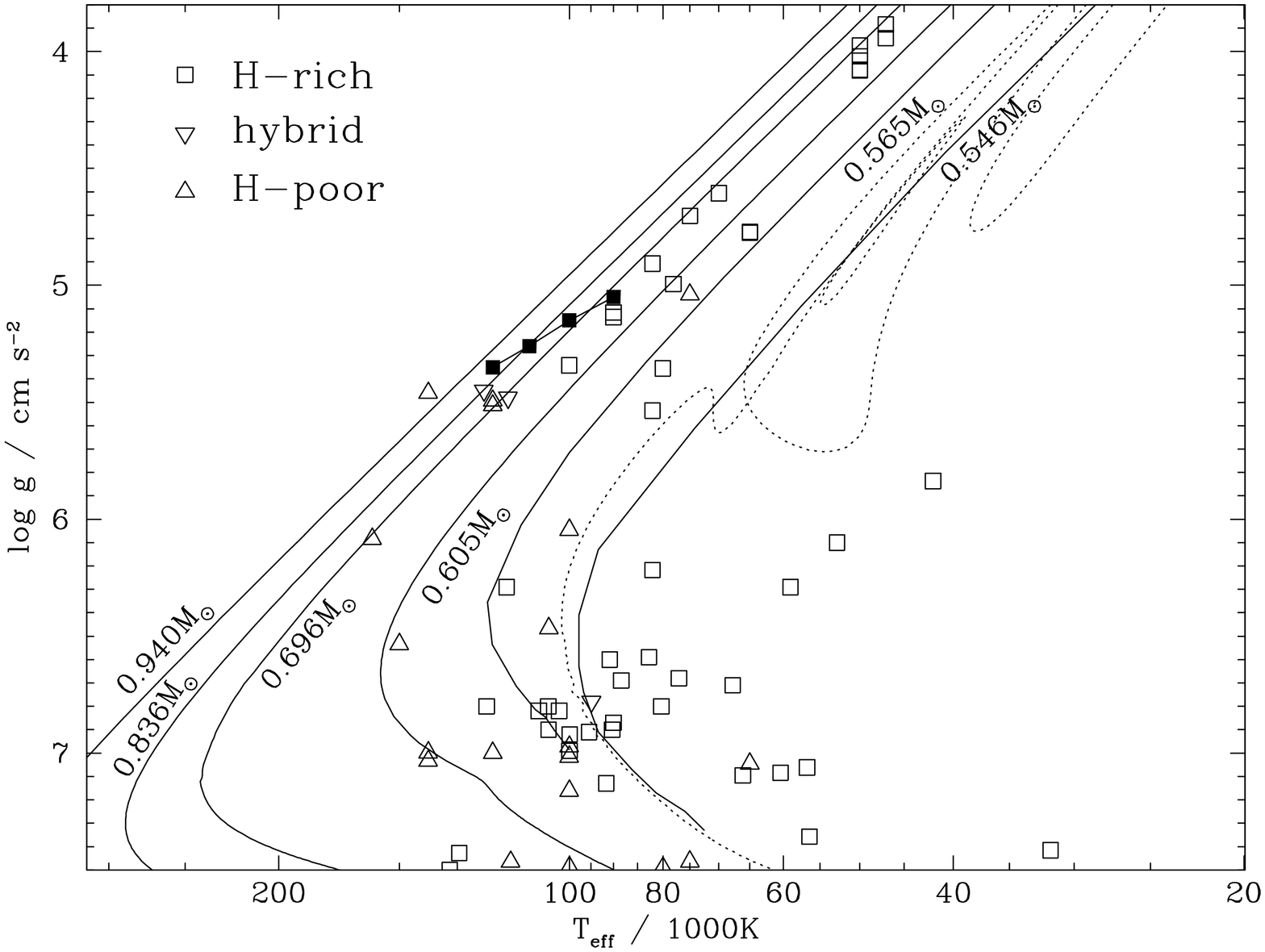}
\caption{Post-AGB evolutionary tracks for different stellar masses
from the grid of \citet{blocker1995} and \citet{sch83}
in the (log $g$, \Teff) plane.
The $0.524M_\odot$ post-AGB track discussed in the text is drawn as dotted line.
The connected filled symbols indicate the positions of the central star
of \png\ according to the solutions ranging
from 90000\,K to 120000\,K. Open symbols are central stars
of planetary nebulae and other post-AGB
objects taken from the literature
\protect\citep{napw1999}}
\label{tefflogg}
\end{figure}

Below 90000\,K and above 120000\,K the quality of the Balmer line
fits degrades significantly.  The analysis of the nebular spectrum
of \png\ indicates a central star temperature of at least
100,000\,K (Tovmassian et al. 2001, Richer et al. 2002).  Taking
into account the detection of the [\ion{Ne}{5}]~3426\,\AA\ line
(Jacoby et al. 2002, Stasi\'nska et al in preparation) an even
higher temperature is required, unless an additional source of EUV
photons is present in this system.  These combined constraints are
best fulfilled by the solution with $\Teff =120000$\,K and $\log g
= 5.35$. The atmosphere model fits to the portions of spectra with
largest RV shifts are presented in Fig~\ref{balmerfit}. The formal
fit uncertainty in $\log g$ for a given value of $\Teff$ is
0.07\,dex. As discussed in \cite{napw1999}, the formal errors
computed by the fit procedures are usually too optimistic. Thus we
made an estimate of the external error as described in
\cite{napw1999} and derived an error limit of 0.22\,dex.
The position of \png in the $\log g, \Teff$ plane is shown in
Fig.~\ref{tefflogg}. Although the helium lines are weak and almost
completely filled in by the nebular emission lines, we could
derive an estimate of the helium abundance (Fig.~\ref{balmerfit}).
The fit yielded $\log n_{\mathrm{He}}/n_{\mathrm{H}} = -1.0$,
i.e.,\ close to the value derived for the nebula. The error of the
abundance measurement can be estimated from the scatter between
the individual spectra, which amounts to 0.2\,dex.

\begin{figure}
\centering \resizebox{\hsize}{!}{\includegraphics[bb=25 0 580 760,
angle=-90, clip]{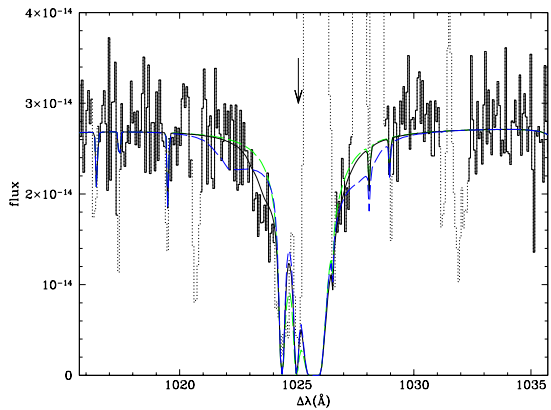}} \caption{Model atmosphere fit to the
Ly\,$\beta$ line of \png. The histogram is the observed all photon
spectrum.  Only those parts shown with a continuous line were used
to fit both the stellar profile and determine the interstellar
absorption. The bold, solid line shows the best fit to the stellar
absorption including all of the interstellar contributions and
smearing effects (where $K_1$ is the radial velocity of the
ionizing star $= 450\,$km/s). The long dashed lines are the
spectra resulting for $K_1 = 0\,$km/s (green) and 900\,km/s
(blue). The arrow indicates the center of the photospheric
Ly$\beta$ line. } \label{f:Lyfit}
\end{figure}

The Ly$\beta$ line from the FUSE spectra is sensitive to gravity
as well. However, the total exposure time of the FUSE spectrum
amounts to 8.5\,hours, distributed over 11.7 hours. This is much
longer than any plausible orbital period (Sect 4.4) and it is
reasonable to assume that we have approximately equal coverage of
all orbital phases. This means that orbital smearing is the
dominant broadening mechanism for the observed Ly$\beta$ line,
even more important than the pressure broadening of the
photospheric absorption line. We adopted the atmospheric
parameters derived above ($\Teff = 120,000$,K and $\log g = 5.35$)
for the fit shown in Fig.~\ref{f:Lyfit}.

The photospheric Ly\,$\beta$ line is
contaminated by interstellar absorption that we must take it into
account. In addition to the discussion in Section 3.1 the column
density of the atomic interstellar hydrogen can be determined from
the red wing of the Ly\,$\beta$ line which is most sensitive to
the interstellar absorption (cf. Fig.~\ref{f:Lyfit}, note the gap
between the terrestrial airglow lines of H~{\sc i} and O~{\sc i}).
Best agreement is achieved with a column density $N_{\mathrm{H}} =
3 \times 10^{20}\mathrm{cm}^{-2}$. Acceptable fits are derived for
the range $2 \times 10^{20}\mathrm{cm}^{-2} \le N_{\mathrm{H~I}}
\le 5 \times 10^{20}\mathrm{cm}^{-2}$. This coincides perfectly
with estimates of $N_{\mathrm{H}}$ from other wavelength regions
(Section 3.1).

As mentioned above the line widths depend upon the assumed
amplitude of the orbital motion, $K_1$.  We obtained the best fit
for a velocity amplitude of 450\,km/s. Acceptable fits could be
produced for velocity amplitudes in the range from 250\,km/s to
600\,km/s. Thus the result from the fit of the Ly$\beta$ line is
in agreement with the orbital solutions discussed in the next
section.

\subsection{Investigating the periodicity of the radial variations}

As mentioned above, {\sc fitsb2} is not only able to derive the
stellar parameters, but can measure RVs as well. We measured RVs
for all CFHT spectra with temperature and gravity now fixed at the
value $\Teff = 120000$\,K and $\log g = 5.35$ derived above. The
Balmer lines $H\gamma$ to H9 and the two \ion{He}{2} lines were
used for the RV determination. The results are summarized in
Table~\ref{rvtbl}. Formal errors provided by the fit procedure are
relatively small ($\approx$15\,km/s).
We added to these uncertainties an additional 10\,km/s to account
for the remaining uncertainty in the wavelength scale.
The RV measurements for the conjunction spectra 471 and 472 are
problematic, because the RV fits basically rely on the invisibility
of the photospheric Balmer lines. We estimated an RV error of
60\,km/s from the scatter of the results derived using different methods.

\begin{figure}[t]
\includegraphics[angle=-90,width=78mm]{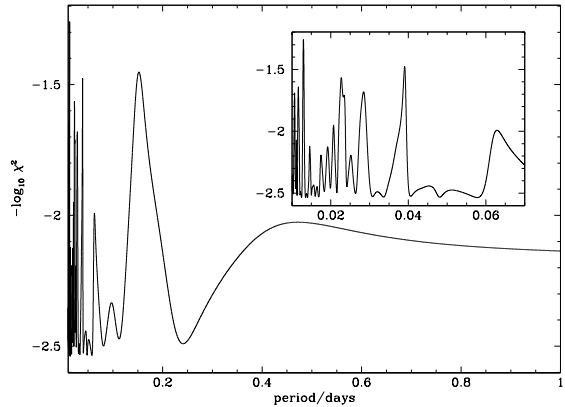}
\caption{The power spectrum
for the {\sc FITSB2} RV measurements from
Table~1. The inset gives a more detailed view for shorter periods.
Periods below 0.01\,d ($\approx$15\,min) can be ruled out, because
these periods would be much smaller than our exposure time.}
\label{power}
\end{figure}

\begin{figure}[t]
\includegraphics[width=81mm]{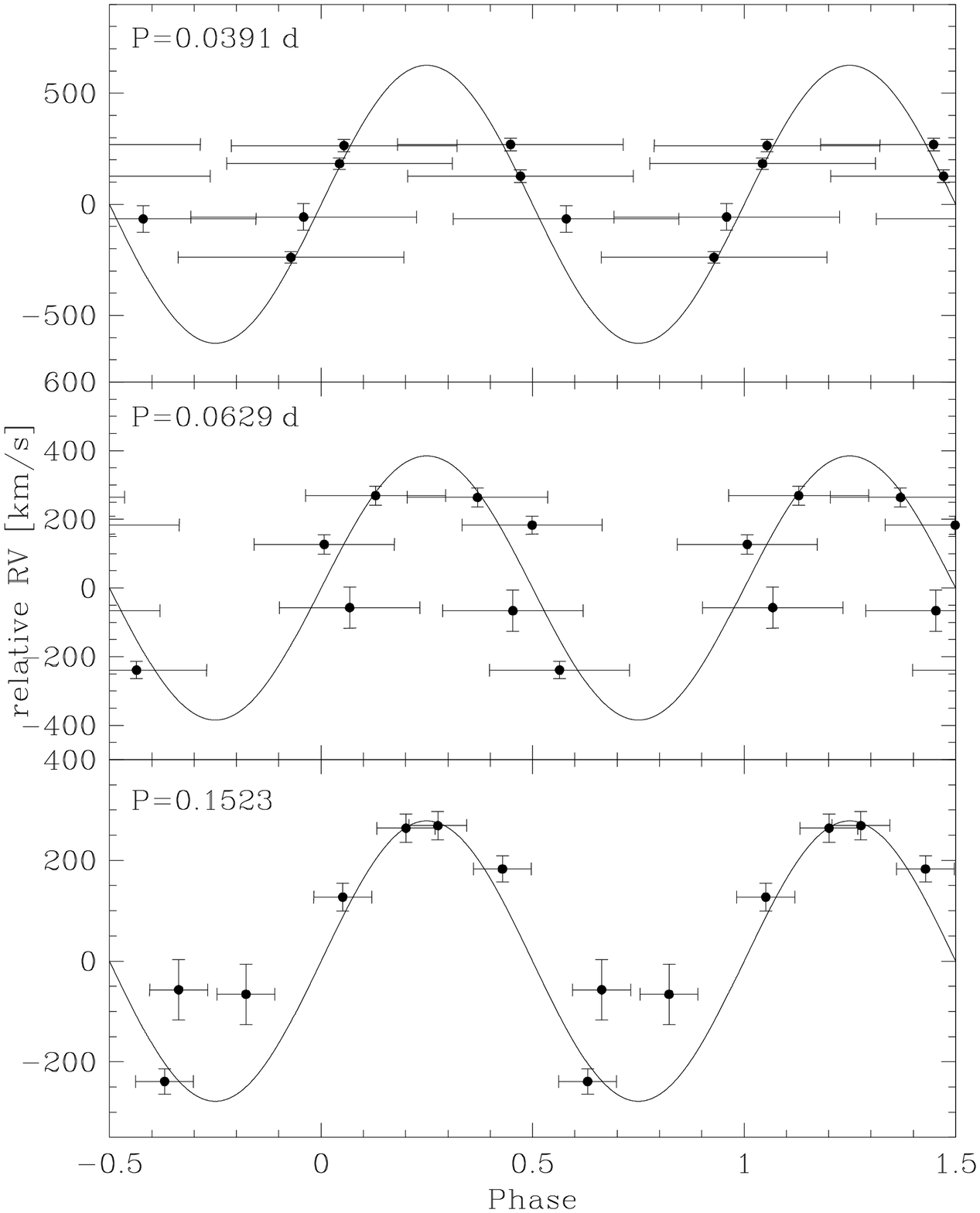}
\caption{Radial velocity curves for three possible orbital
periods. The error bars in the phase coordinate correspond to the duration
of the individual exposures.}
\label{rvcurves}
\end{figure}

The RV variation has a clearly periodic nature. However, the
sparseness of data does not allow an unambiguous determination of
the orbital period. Fig.~\ref{power} shows a ``power spectrum''
which was produced by fitting RV sine-like curves for a range of
periods. The resulting fit quality, measured by $\chi^2$, is
plotted in Fig.~\ref{power}. Since the number of data points is
low and the measurements are difficult due to the strong nebular
emission lines, we held the system velocity fixed at the value RV
= $-$193\,km/s derived by \citet{richeretal2003}. The best fit
quality is obtained for a period of 0\fd 150 (3\fh 6). The
corresponding RV curve is shown in the bottom panel of
Fig.~\ref{rvcurves}. For this solution, the two RV measurements
close to conjunction (phase 0.75) are problematic, because they
lie in a ``forbidden'' area of this diagram. These points are less
problematic for the solutions with shorter periods of 0\fd 039
(0\fh 94) and 0\fd 063 (1\fh 5), presented in the top and middle
panels of Fig.~\ref{rvcurves}, respectively. Orbital parameters
are detailed in Table~\ref{orbits}. The amplitudes from the best
fit solutions for 0\fd 039\,d and 0\fd 063 are quite high. We
calculated alternative (``minimal'') solutions by assuming that
the orbital amplitudes correspond to the maximum observed RV
shifts (after correction for orbital smearing; see below).  A
solution with a period significantly longer than 4\,h can be ruled
out.  Apart from the formal $\chi^2$ values shown in
Fig.~\ref{power}, simply consider the large RV shift between
spectra 457 and 466: 470\,km/s in just 2\fh 3!

For the two shortest period solutions found (0\fd 063 and 0\fd
039), note that the exposure times of 30\,min extend over a large
range of phases. Thus, smearing due to the orbital RV variations
must be taken into account as a major problem.  This could cast
doubt upon whether any absorption line features should be
observable at all. We made simple simulations of this effect and
found that, although the line profiles are indeed seriously
distorted, line profiles similar to those observed can still be
produced, but that the measured RV values are severe
underestimates of the real RVs.
The difference between the real and measured RVs are smallest
for the exposures centered on the quadrature phases of largest RV
shift. We calculated approximate corrections for these phases by
measuring the displacement of the line cores in our simulated
spectra. Results are 124\,km/s for the 0\fd 039 orbit and 59\,km/s
for 0\fd 062. The correction for 0\fd 151 is negligible within
uncertainties (10km/s).
We can conclude that the orbital period cannot be longer than
4\,h, but a time series of spectra with much shorter exposure
times are needed for an unambiguous period determination.

It has long been speculated that many planetary nebulae have
binary nuclei. There are models that invoke binarity of the central
object in order to explain the
morphology of certain types of planetary nebulae
\citep{soker2002}. Evolutionary models of close binaries require
that a fraction of the systems going through the common envelope
phase form a planetary nebula \citep{yutl1993,IT1989}.  However,
the number of observationally-confirmed binary cores in planetary
nebulae is relatively small.  Nonetheless, there are
reasons to believe that many more of them are indeed binaries
\citep{demarcoetal2003}, though many probably have relatively long
periods. The discovery that the central star of \png\ is a close
binary is thus very important in this context.

\begin{table*}[t]
\caption{Orbital solutions for a $0.55M_\odot$ primary and the
three possible periods. The set of parameters on the left
corresponds to the fit results displayed in Fig.~\ref{rvcurves}.
For the set of parameters on the right, we assumed that the
orbital velocity corresponds to the maximum of the  measured RV
shifts, corrected for the effect of orbital smearing.  The
subscripts 1 and 2 correspond to parameters for the ionizing and
companion stars, respectively.} \vspace{3mm} \label{orbits}
\begin{tabular}{lrllllrllll}
\tableline\tableline
&\multicolumn{5}{c}{Best fit solution} &\multicolumn{5}{c}{``Minimal''
solution}\\
$P$  &$v_{\mathrm{orb}}$ &$M_2\sin i$  &$a$ &$R_1^{\mathrm{Roche}}$
     &$R_2^{\mathrm{Roche}}$   &$v_{\mathrm{orb}}$ &$M_2\sin i$
     &$a$ &$R_1^{\mathrm{Roche}}$   &$R_2^{\mathrm{Roche}}$ \\
(d)  &(km/s)     &($M_\odot$) &($R_\odot$) &($R_\odot$) &($R_\odot$)
     &(km/s)     &($M_\odot$) &($R_\odot$) &($R_\odot$) &($R_\odot$)  \\
\tableline
0.039  &590  &1.55  &0.63  &0.19  &0.30  &390 &0.73 &0.54 &0.19 &0.22\\
0.063  &380  &0.92  &0.77  &0.26  &0.33  &320 &0.69 &0.73 &0.26 &0.29 \\
0.151  &265  &0.82  &1.35  &0.47  &0.56  \\
\tableline\tableline
\end{tabular}
\end{table*}

\section{The extinction of \png\ }

With our FUSE data, we are able to further address the question of the
reddening of PN\,G\,135.9 +55.9.  In our previous studies, the reddening was derived
from a comparison of the observed hydrogen and helium emission line
ratios with those given by recombination theory.  \citet{richeretal2002}
noted  that the reddening of \png\ was modest. The H$\gamma$/H$\beta$
and H$\delta$/H$\beta$ line ratios gave an $E(B-V)$ of 0.3 -- 0.35 mag
(the H$\alpha$/H$\beta$ ratio was apparently variable and indicated
even lower reddening). The He~{\sc ii} $\lambda$ 5412/4686 ratio
implied a negative reddening.

\begin{figure*}
{\includegraphics[width=168mm,
bb=20 100 580 750, clip]{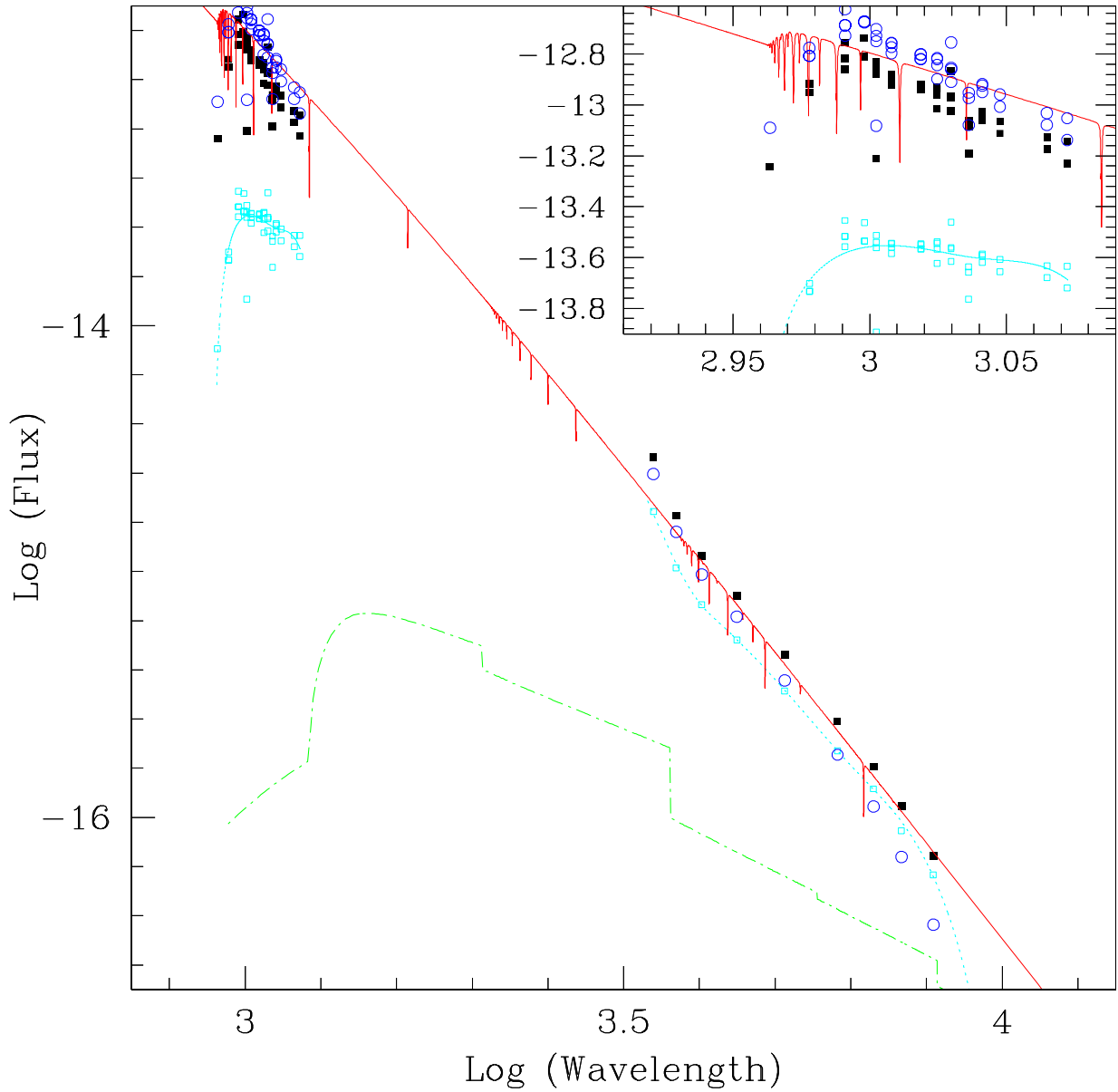}}
\caption{The flux distribution of \png\ from far UV to near IR.
The open squares are the observed fluxes. The dashed lines are
continuum fits to the observed data.  Fluxes corrected for interstellar
extinction are represented by filled squares ($R_{\rm V}$ = 3.1,
$E(B-V)$ = 0.045 mag). The dot-dash line is the expected contribution
from the nebula (see text).  The open circles are the
extinction-corrected data ($R_{\rm V}$ = 2.3) after subtraction of
the nebular component.
The central star model with 120\,000\,K and log\,$ g = 5.35$ is
presented as a solid line. The model is normalized to the raw flux
observed at 4\,000\,\AA.} \label{uvbb}
\end{figure*}

From the FUSE data, we can estimate the extinction suffered by the
object  by fitting the flux distribution of \png\ from the far UV
to the near IR. This is shown in Fig. \ref{uvbb}. In this figure,
the observed energy distribution of \png\ from the far UV to the
near IR is represented by open  squares. The fluxes were measured
in  narrow bands free of any significant lines ($\approx
3-4\,$\AA\ in the far UV and $\approx 20\,$\AA\ in the optical).
The dashed lines are continuum fits to the spectrophotometric
data. The solid line is the model of the central star described
above for a temperature of 120\,000\,K and log $g$ =5.35. The
slope of the observed UV spectrum (1\,000 -- 1\,200\,\AA) is
consistent with the model, but shifted downwards due to the
interstellar extinction. If we correct for interstellar reddening
using the tables from \citep{fitz} and assume the standard value
$R_{\rm V}$ = 3.1 and $E(B-V)$ = 0.03 mag \citep{jacobyetal2002},
we effectively recover $\approx 0.45$\,mag of flux, but this is
not enough to compensate all losses in the UV. Increasing $E(B-V)$
to 0.045 mag yields a better fit overall, but, at that point, the
optical fluxes show some excess in comparison to the model, while
the UV flux is still somewhat short of it. If we also take into
account the nebular continuum at a temperature of 30\,000\,K from
our best fit photoionization models of the nebula \citep[][Sect.
8]{richeretal2002}, which contributes primarily in the optical
range (dash-dot line in Fig. \ref{uvbb}), then the fit is improved (filled
black squares). Much better results are obtained using $R_{\rm V}$
= 2.3 and $E(B-V)$ = 0.045 mag (open circles). This could mean
that the dust towards \png\ is non-canonical.  This may indeed be
the case since the line of sight toward \png\ is not necessarily
characterized by the canonical value $R_V$ = 3.1
\citep[e.g.,][]{barb2001}. Moreover, the far UV part of
Fitzpatrick's reddening curve is actually an extrapolation of near
UV data into the FUSE range.

Another way to estimate the reddening is to use the H~{\sc i}
column density derived from our FUSE spectrum and convert it into
a reddening using the relation between $E(B-V)$ and
$N_\mathrm{H~I}$ of \citet{buhe1978}. For $N_{\mathrm{H~I}}$ = $3
\times 10^{20}$, we find $E(B-V)$ = 0.01 mag.

These values are somewhat smaller than the reddening derived by
\citet{richeretal2002} from the Balmer decrement. Those reddenings
were derived in the usual way, comparing the observed emission
line ratios with the theoretical ones and assuming emission under
case B at a nebular temperature of 30\,000\,K. However, no account
was taken of the underlying stellar absorption. When this effect
is properly taken into account (Stasi\'nska et al. in
preparation), the reddening derived from the Balmer decrement is
compatible with the values obtained from our FUSE spectrum.

Thus  we confirm that the reddening, and hence extinction, of
\png\ is indeed low. In the following, we  adopt a value of
$E(B-V)$ = 0.04 mag, but none of our conclusions would be
significantly affected should the real value be slightly
different.

\section{The distance and age of \png\ }

Having an estimate of the stellar gravity and a measurement of the
apparent magnitude of the central star, we can estimate the
distance $d$ to the object using the expression:

$$d =1.11 (M_{\star} F_{\rm V} 10^{0.4 m_{\rm V}}/g)^{0.5} $$ where $d$ is
in kiloparsecs, $M_{\star}$ is the stellar mass in solar units,
$F_{\rm V}$ is the model stellar flux in the $V$ band in $10^8$
erg cm$^{-2}$ s$^{-1}$ \AA$^{-1}$, and $m_{\rm V}$ is the observed
stellar magnitude corrected for nebular contribution and for
extinction. This method to evaluate the distances to PNe has been
discussed extensively by \cite{napw2001} and there are strong
arguments that such spectroscopic distances are reliable (within
the uncertainties). To evaluate the distance to our object, we
take the observed magnitude $m_V = 17.9$\,mag from Richer et
al.(2002), assume a reddening of $E(B-V)$ = 0.04\,mag, and apply a
correction of 0.15\,mag for the nebular continuum contribution, as
derived from photoionization models described in Richer et al.
(2002). The stellar flux for the model with $\Teff =120000$~K and
$\log g = 5.35$ is $7.07\cdot 10^8$ erg cm$^{-2}$ s$^{-1}$
\AA$^{-1}$. Adopting a stellar mass of 0.55\msun, we find a
distance to \png\ of 17.8~kpc. Had we adopted a stellar mass of
0.88\msun, as indicated by Fig. 4 and listed in Table 2, we would
have found a distance of 22.6~kpc. Assuming that the stellar
temperature is only 90000~K and taking the values of $\log g$ and
$M_{\star}$ from Table 2, we  find $d$ = 24.2~kpc, while assuming
a stellar temperature of 150000~K, we find $d$ = 20.6~kpc. The
largest contribution to the error budget is  the uncertainty in
log $g$ for a fixed \Teff, which amounts to an uncertainty of
5\,kpc in the distance.

In any case, our present estimates of the distance are consistent
with \png\ being located in the Galactic halo. Note that the
values of the distance obtained in this way are roughly consistent
with the ones obtained previously
\citep{tovmassianetal2001,richeretal2002}, which were based upon
nebular properties (flux and dimensions), whereas the
present determination relies \emph{only} on the fit of the stellar
spectrum. This indicates that we are progressively getting a
coherent picture of the  nebula.

\begin{figure}[t]
 \includegraphics[width=66mm, angle=-90]{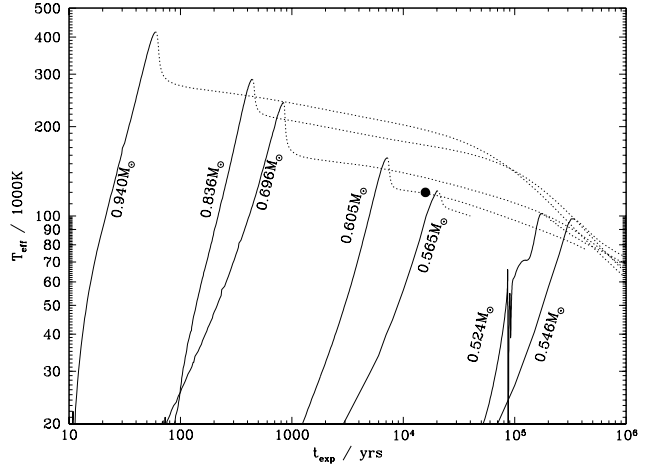}
\caption{Post-AGB evolutionary tracks for different stellar masses
interpolated from grid of \citet{blocker1995,sch83} in the (T$_{\rm
eff}$, t$_{\rm exp}$) plane. Tracks are plotted with solid/dotted
lines for the parts before/after maximum temperature.
The position of \png\ is marked for the
estimated values of T$_{\rm eff}=120\,000$\,K and t$_{\rm
exp}=16\,000$\,yr.}
\label{expage}
\end{figure}

On the other hand, the discovery of the compact binary nature of
the nucleus of \png\ introduces new challenges in interpretation
of the evolutionary path of this unique object. For further
discussion, a rough estimate of the age of the planetary nebula is
useful.  The expansion time, $t_{\rm exp}$, of the planetary
nebula is generally computed as $t_{\rm exp}$ = $R_{\rm
out}$/$v_{\rm exp}$, where $R_{\rm out}$ is the outer radius of
the nebula and $v_{\rm exp}$ is its expansion velocity. This is of
course a simplistic approach, since it is known that the expansion
velocity should vary during the course of the nebular evolution
\citep[see e.g.\ the dynamical models of ][]{vimaga2002} and that
the epoch of the ejection of the nebula does not necessarily
coincide with the evolution of the central star off the AGB. In
the case of \png\ the expansion time measures the time since a
final mass loss episode triggered by common envelope interaction
of the binary stellar core (Sect. 7).  The transition phase from
the AGB to a hot post-AGB was certainly strongly affected by the
binary evolution. Nevertheless, an estimate of the nebula
expansion time provides a valuable constraint, which is especially
important in the context of close binary stellar evolution. We
adopt the nebular size determined by \citet{richeretal2002} and an
expansion velocity of 30 km s$^{-1}$ as found by
\citet{richeretal2003}. For a distance of 17.8~kpc, we find a
nebular expansion time of 15800~yr. Using this rough age estimate,
the position of the object in the (T$_{\rm eff}$, t$_{\rm exp}$)
diagram (Figure \ref{expage}) can be compared with that of model
stellar tracks of various masses \citep{blocker1995,sch83}.  The
position of \png in Fig.~\ref{tefflogg} indicates that the
comparison must be made with the part of the tracks that lie prior
to the temperature maximum (solid lines in Fig.~\ref{expage}). We
find that the position of \png\ is consistent with a mass of the
ionizing star around 0.57\msun.

\section{Discussion: The status of the stellar core of \png}

\subsection{The ionizing star}
As described above, we derived the surface gravity of \png from
fits of the Balmer lines of selected optical spectra. Because of
nebular contamination of the optical spectra, we cannot determine
both the temperature and gravity, so we adopted $\Teff =
120000$\,K based upon the analysis of the nebula (Stasi\'nska et
al. in preparation). The resulting gravity is $\log g = 5.35$.
From a comparison with evolutionary tracks of H-burning central
stars of planetary nebulae (Fig.~\ref{tefflogg}) we estimated a
mass of $0.88M_\odot$.  This value is at variance with the
expected properties of a Population~II central star. Three lines
of evidence point towards a Population~II nature of \png: the
large height above the Galactic plane ($z=14.8$\,kpc for $d=17.8$\,kpc),
the high radial velocity (RV $=-193$\,km/s), and the very low metal
abundances derived from the analysis of the planetary nebula and the
far-UV spectrum of the central star (Stasi\'nska et al., in
preparation).

Known post-AGB stars of Population~II have low masses
\citep[$\approx 0.55M_\odot$; e.g.,][]{Moehler98}, which is
explained by the small mass of their progenitors. One could argue
that problems caused by the orbital motion and the contamination
by the nebula emission lines cause a systematic shift of the fit
results. However, from a comparison of observed and synthetic
spectra, we conclude that a gravity much higher than that given by
our fits can be ruled out, because the resulting lines would be
much broader than observed. Our estimate of the effective
temperature is based upon the analysis of the nebular spectrum.
If an additional source of hard ionizing photons were present in
this system, (e.g.\ resulting from interaction between the components
of the binary system), the nebular spectrum would actually
overestimate the temperature of the ionizing star.  Lower
temperatures would result in lower mass estimates (Table~\ref{tgfit})
in better agreement with the expected mass of a Population~II central
star.

Due to the obvious Population~II nature of \png, in the subsequent
discussion we shall adopt a mass of $0.55M_\odot$ for the ionizing
star. A low mass is also supported by the relatively large
kinematic age of the nebula. If the mass of the ionizing star were
really as high as the estimates derived from Fig. 4, e.g.,
$0.9\,M_\odot$, the time interval between the tip of the AGB and
the hottest point of the post-AGB evolution would be of order of
only 100 years, much shorter than the estimated nebular expansion
age of 16000 years. The observed nebular age is in much better
agreement with a mass of $\approx 0.57M_\odot$ (or lower), for
which transition times are $\approx$20000\,yrs (or longer). Note
that the common envelope phase very likely abbreviated the
post-AGB evolution.

How might we explain the apparent discrepancy between the observed
and expected parameters of the ionizing star? The progenitor
almost certainly went through a common envelope phase, which drove
it out of thermal equilibrium.
In this case, the relations used to derive the stellar mass
from log $g$ and \Teff\ do not apply. However, relaxation happens
on the thermal time scale of the post-AGB envelope, which is given
by
\begin{displaymath}
t_{\mathrm{th}} = G \frac{M\,M_{\mathrm{env}}}{R\,L} \: ,
\end{displaymath}
where $G$ is the gravitational constant, $M$ the mass of the
central star, $M_{\mathrm{env}}$ the mass of the stellar envelope,
$R$ the stellar radius, and $L$ the stellar luminosity.  For the
derived stellar parameters and an adopted mass for the post-AGB
envelope of $M_{\mathrm{env}}=10^{-2}M_\odot$, the resulting
thermal time scale amounts to only $\approx$30 years. Since the
real envelope mass is probably smaller \citep[cf.\ Fig.~3 of
][]{sch83}, this can be considered an upper limit. The short
relaxation time makes it unlikely that we observe the star
precisely in this stage.

A more plausible explanation is that the post-AGB star experienced
a late thermal pulse (shell helium flash) after emerging from the
common envelope phase. Stars suffering a late thermal pulse evolve
back to cooler temperatures and then replicate the previous
post-AGB evolution \citep{IKT83, sch83, blocker1995}.  The
evolutionary speed and luminosity during this second phase can be
quite different from the ones in the first phase. As an
illustration, we plotted in Fig.~\ref{tefflogg} the track computed
by \citet{blocker1995} for a $0.52M_\odot$ post-AGB star, which
experiences two late thermal pulses. The resulting track spans a
wide range of gravities and luminosities, whose details depend
upon the precise mass and thermal pulse phase.  Most importantly,
the time scales for this kind of evolution are thousands to tens
of thousands of years, making it much more likely to catch a star
in one of these phases.

\subsection{The companion star}

The short orbital period and large semiamplitude of the RV that we
detect imply that the other component of the binary system should
be a rather massive star. The mass function can be used to derive
lower limits to the mass of the unseen companion, if the mass of
the visible star is known (Table~\ref{orbits}). In the case of the
3\fh 6 solution we derive a minimum companion mass of $0.82
M_\odot$ for a $0.55M_\odot$ ionizing star ($1.01M_\odot$ for a
$0.88M_\odot$ ionizing star). The best fit orbits for
the shorter period solutions indicate even higher companion
masses. However, the ``minimal'' solutions yield lower limits of
$\approx$$0.70M_\odot$. Thus the companion must be a relatively
massive star.

Note that the system is physically small. The separation between
both stars amounts to only $1.35 R_\odot$ for the 3\fh 6 solution
and the $0.55M_\odot$ ionizing star (Table~\ref{orbits}). The
Roche lobe of the companion has a size of $0.56R_\odot$. A
metal-poor $0.6M_\odot$ main sequence star has a radius of
$0.55R_\odot$ \citep{Baraffeetal1997},  almost filling its Roche
lobe if the period is 3.6\,h and overfilling it for the other
solutions. However, the orbital solutions require more massive
companions that have even larger radii and would therefore not fit
into their Roche lobes, so a main sequence companion appears
unlikely. Further evidence against a main sequence companion comes
from the the binary system BE\,UMa. It consists of a hot post-AGB
star with parameters similar to that of \png
\citep{Liebertetal1995} and a K dwarf companion. However, this
system has a much wider orbit ($P=2\fd 29$) than \png, yet it
still shows a spectacular reflection effect producing a forest of
RV variable emission lines from the heated surface of the
companion \citep{FerJam1994}. The absence of these lines in our
spectra is another strong argument in favor of a compact
companion, i.e., a white dwarf or a neutron star. A photometric
monitoring campaign to search for variability  should be able to
provide final evidence against (or for) a main sequence companion.
Note that the ionizing star fills a significant fraction of its
Roche lobe as well (Tables~\ref{atmmod} and \ref{orbits}). The
$\Teff=120000$\,K, $\log g = 5.35$ solution corresponds to a
radius of $0.26R_\odot$ compared to a Roche lobe of $0.47R_\odot$
in the 3.6\,h orbit. According to the temperature-radius relation
from the $0.546M_\odot$ track of \citet{sch83} the ionizing star
completely filled its Roche lobe when it had a temperature of
50000\,K. This represents the minimum temperature of the ionizing
star for the stellar core of \png to become a detached system.
Model predictions indicate that the central star fills or
overfills its Roche lobe for the 1.6\,h and 0.94\,h solutions,
respectively.

In the standard picture of the common envelope phase the companion
spirals in until enough orbital energy is released to eject the
envelope \citep{Web1984}. For a system with a companion as massive as
in the case of \png, one would expect that only a moderate spiral in
would be necessary, producing a relatively wide system. We used
standard assumptions for the common envelope phase \citep[formula A.14
of][with $\alpha_{\mathrm{CE}}\lambda =2$]{Nel2001} to estimate the
initial separation of the \png\ system before the final common
envelope. For the 3\fh 6 orbit we calculated an initial separation of
only $3.6R_\odot$ (initial/final mass: $0.85M_\odot$/$0.55M_\odot$,
companion mass: $0.8M_\odot$).  This is an impossible value, because
this would indicate that interaction started shortly after the main
sequence phase, preventing the star from becoming the central star of
a planetary nebula. Thus it is difficult to explain the formation of
the \png\ system with standard assumptions. On the other hand this
means that \png\ could provide an important test of our understanding of
common envelope evolution.

If the companion is a compact star, the binary will merge in less
than 1\,Gyr due to the loss of energy and angular momentum caused by
gravitational wave radiation.  Our estimates from the mass
function show that the companion should be rather massive and it is
quite possible that the sum of the masses of this system exceeds
the Chandrasekhar limit for white dwarfs ($1.4M_\odot$). Such
systems are proposed as progenitors of SN\,Ia
\citep{Web1984,IT1989}.  The SPY project (Supernovae Ia Progenitor
surveY) \citet{NCD2001,NCD2003} performed a radial velocity survey
of white dwarfs for potential SN\,Ia progenitors. To date, only
one possible candidate system was detected \citep{NCD2003},
illustrating how rare these systems are. If it can be shown that
the core of \png\ is massive enough to qualify as SN\,Ia
progenitor, this would provide the bonus that additional
information about the formation of such systems is available from
the chemistry and morphology of the planetary nebula.

\section{Conclusions}

We have presented new optical and far-UV spectra of the central
star of the very metal-poor planetary nebula \png.  The
photospheric Balmer lines visible in the optical spectra enabled
us to determine the surface gravity from a model atmosphere
analysis. For a temperature of $\Teff=120000$\,K (as indicated by
the nebular emission lines), we derived a gravity of $\log g =
5.35$ and a photospheric helium abundance $\log
n_{\mathrm{He}}/n_{\mathrm{H}}=-1.0$. Our model atmosphere
spectrum also fits the UV Ly$\beta$ line profile observed in the
FUSE spectrum satisfactorily given that the Ly$\beta$ line is
contaminated by interstellar lines and suffers from severe
smearing effects due to the orbital RV variations.

Comparing the above values of $\Teff$ and $\log g$ with those from
evolutionary tracks in the $\log g$-$\Teff$ plane yields a mass of
$0.88 M_\odot$ for the ionizing star, if one adopts standard
H-burning post-AGB stellar tracks.
This result is in contrast with the Population~II nature of this
object, indicated by its metallicity (Tovmassian et al. 2001,
Richer et al 2002, Jacoby et al. 2002, Stasi\'nska et al., in
preparation), location in the galactic halo, and its large radial
velocity \citep{richeretal2003}. We argue that the real mass
of the ionizing star of \png\ is close to the typical masses of
Population~II post-AGB stars: $\approx 0.55M_\odot$. Since
the stellar core of \png\ is a close binary the formation of the
planetary nebula is likely to have happened during a phase of
common envelope evolution.

A comparison of a model spectrum calculated for appropriate
stellar parameters and the observed FUSE and optical fluxes
indicates a reddening of $E_{B-V}=0.04$\,mag. This relatively
small value is in agreement with the expectations for an object at
high galactic latitudes ($b=+56^\circ$).  This low reddening is
also in agreement with the estimates the interstellar atomic and
molecular hydrogen column densities towards \png\ based upon the
FUSE observations.  Allowing for extinction,
we obtain a distance for \png\ from a comparison of its measured
flux in the $V$ band and model atmosphere fluxes. The result is
$\approx$18\,kpc, which places \png\ 15\,kpc above the galactic
plane. The kinematical age of the nebula is $\approx$16000 years,
in rough agreement with expectations if the central star mass is
of order $0.55M_\odot$.  This expansion age argues against a more
massive central star, e.g., $0.9M_\odot$, since the expected
post-AGB age of such a star is only 100\,years.

The optical spectra reveal large radial velocity variations
of the stellar absorption lines, obviously caused by orbital
movement in a close binary system. Since the observations were
not designed for the measurement of a binary orbit, the phase
coverage and the wavelength calibration are not optimal. However,
we show that the orbital period must be less than 4 hours (our
three best solutions are 0\fh 94, 1\fh 5, and 3\fh 6 hours) and
the minimum (half-)amplitude of the radial velocity amounts to
250\,km/s. The companion remains invisible in the spectra during
all orbital phases. Our preliminary orbital solutions indicate a
companion mass of $\approx 0.9M_\odot$. A solid lower limit is
$0.6M_\odot$. It is very unlikely that the companion is a main
sequence star, because we would then expect a strong reflection
effect producing very strong emission lines from the irradiated
hemisphere of the main sequence star. That these are not observed
in the optical spectra indicates that the companion is most
probably a white dwarf or a neutron star.

Due to the short orbital period (four hours or less) the system
will merge in less than one Gyr, if the companion is a compact
star. If a companion mass exceeding $0.85M_\odot$ can be
confirmed, the total mass of the stellar binary core would exceed
the Chandrasekhar limit and this system would qualify as a
potential progenitor of a type Ia supernova.
New observations, with shorter exposure times and longer time
coverage, will be needed in order to establish a definitive
orbital period and other parameters for the binary system.

\begin{acknowledgements}

G.T.\ acknowledges CNRS/CONACyT grant and hospitality of
Observatory de Meudon during his visit. GS and MR acknowledge
CONACYT grant 37214-E. R.N.\ acknowledges support by an PPARC
Advanced Fellowship. We thank Martine Mouchet for help with the
interstellar molecular hydrogen lines and Slawomir G\'orny for
extending the grid of interpolated post-AGB evolutionary tracks.
R.N.\ thanks Philipp Podsiadlowski and Martin Beers for useful
comments on this system. Special thanks to Klaus Schenker for many
helpful suggestions and discussions. This paper is based upon
observations made with the NASA-CNES-CSA Far Ultraviolet
Spectroscopic Explorer. FUSE is operated for NASA by the Johns
Hopkins University under NASA contract NAS5-32985.

\end{acknowledgements}



\end{document}